# RNN-based speech synthesis using a continuous sinusoidal model

Mohammed Salah Al-Radhi
Department of Telecommunications and Media Informatics
Budapest University of Technology and Economics, Budapest, Hungary
malradhi@tmit.bme.hu

Tamás Gábor Csapó
Department of Telecommunications and Media Informatics
Budapest University of Technology and Economics, Budapest, Hungary
csapot@tmit.bme.hu

Géza Németh
Department of Telecommunications and Media Informatics
Budapest University of Technology and Economics, Budapest, Hungary
nemeth@tmit.bme.hu

*Abstract*—Recently in statistical parametric speech synthesis, we proposed a continuous sinusoidal model (CSM) using continuous F0 (contF0) in combination with Maximum Voiced Frequency (MVF), which was successfully giving state-of-the-art vocoders performance (e.g. similar to STRAIGHT) in synthesized speech. In this paper, we address the use of sequence-to-sequence modeling with recurrent neural networks (RNNs). Bidirectional long short-term memory (Bi-LSTM) is investigated and applied using our CSM to model contF0, MVF, and Mel-Generalized Cepstrum (MGC) for more natural sounding synthesized speech. For refining the output of the contF0 estimation, post-processing based on time-warping approach is applied to reduce the unwanted voiced component of the unvoiced speech sounds, resulting in an enhanced contF0 track. The overall conclusion is covered by objective evaluation and subjective listening test, showing that the proposed framework provides satisfactory results in terms of naturalness and intelligibility, and is comparable to the high-quality WORLD model based RNNs.

## I. INTRODUCTION

Statistical parametric speech synthesis (SPSS) is an approach that aims to make the quality of synthetic speech to be as good as recorded speech [1]. Although a number of contextual factors affect the naturalness of the speech, such as phonetic and linguistic features, the advantages of flexibility to change the voice characteristics, robustness, and compact footprint make SPSS to always stay in the attractive attention of many researchers. These benefits have resulted that the SPSS based on hidden Markov models (HMMs) has become substantially popular during the last decade [2]. Such a parametric HMM approach, decision-tree clustering and context-dependent phonemes are the most commonly used to represent distributions between acoustic and linguistic features [2] [3]. Even though these models can increase the robustness of parameter estimation and generation, the detailed characteristics of the reconstructed spectral envelopes are usually over-smoothed due to the statistical processing in HMM training tend to make the quality of the synthetic speech muffled [4]. There have been, and are still, some successful attempts based on HMMs to alleviate the over-smoothing problem, including trajectory model [5] and global variance [6]. However, the computational complexity (number of parameters and memory) of refined models usually increases gradually.

Recently, an interesting approach based on neural networks have achieved remarkable success in various machine learning tasks [7] and have been also successfully applied to the modeling of speech signals, such as speech synthesis [8], speech enhancement [9], and voice conversion [10]. Deep neural networks, with more than one layer of hidden units between its input and output layers, offer improvements over HMMs including the ability to model long-span frames, highly complex and strongly correlated features [11], hierarchical structure [12], and multi-task learning [13].

Such approaches have been applied in SPSS, like the feed-forward deep neural network (FF-DNN) that maps a finite input context (often 5 frames) to a finite output, and shows significant performance improvements in vocoding features and acoustic modeling based speech synthesis [14]. Even though these approaches achieve good improvements in speech applications, the accuracy of the synthesized speech is still degraded compared to the natural voice for two main reasons. First, FF-DNN still lack the ability to capture long-term dependencies [15] that are spread across time to affect prediction output. Second, the sequential nature of speech is ignored for the reason that each frame is sampled independently in FF-DNN.

In order to alleviate these difficulties, FF-DNN have been replaced by recurrent neural networks (RNNs) - inserting cyclical connections - which allows possibly infinite context. Successful approaches based on RNNs had also been used in the literature to limit the negative impacts on capturing long term dependencies. In [16], a long short-term memory (LSTM) was proposed to use memory cells to retain the past for longer duration. In [17] a bidirectional LSTM (Bi-LSTM) was applied in which there is a feedback to retain previous states, and have shown to outperform FF-DNNs based SPSS [18].

Recently, researchers came up with RNN based analysis/synthesis models such as WaveNet [19], SampleRNN [20], and Tacotron [21]. Those models, so called neural vocoders, have the ability to generate a high-quality synthesized speech from acoustic features without vocoding as an intermediate step. Despite the fact that most of these models outperformed the previously used speech synthesis approaches [22], they needed external modules to extract the linguistic features and some vocoder parameters. Thus, simple and uniform vocoders using RNNs, which would handle all speech sounds and voice qualities in a unified way, are still missing in SPSS. Therefore, we are trying to develop a solution in this paper to achieve higher sound quality with simple vocoder.

This paper is organized as follows: Section II describes the related work. In Section III, we propose the novel idea of continuous sinusoidal model based on RNN. In Section IV, experimental conditions and error metrics are addressed. We report the objective and subjective evaluation results in Section V. Section VI gives the conclusion and discussion of future works.

## II. RELATED WORK

In our earlier work, we proposed a computationally feasible residual-based vocoder [23], using a continuous F0 model [24], and Maximum Voiced Frequency (MVF) [25]. In this method, the voiced excitation consisting of pitch synchronous PCA residual frames is low-pass filtered and the unvoiced part is high-pass filtered according to the MVF contour as a cutoff frequency. The approach was especially successful for modelling speech sounds with mixed excitation. However, we noted that the unvoiced sounds are sometimes poor due to the combination of continuous F0 and MVF. In [26], we further control the time structure of the high-frequency noise component by estimating a suitable time True envelope.

In [27], we successfully modelled all vocoder parameters (continuous F0, MVF, and MGC) with FF-DNNs and shown that the FF-DNN have higher naturalness than HMM based text-to-speech. In [28], we extended modeling of our continuous vocoder parameters using RNN, LSTM, BLSTM, and gated recurrent network (GRU) variants. Experimental results demonstrate that using Bi-LSTM systems to train continuous vocoder parameters improves the synthesis performance and outperforms FF-DNN and other recurrent topologies. The advantage of a continuous vocoder in this scenario is that vocoder parameters are simpler to model than conventional vocoders with discontinuous F0.

Previous studies have shown that human voice can be modelled effectively as a sum of sinusoids and has shown the capability of providing high-quality copy synthesis and prosodic modifications [29] [30] [31]. Therefore, in [32] we proposed a continuous sinusoidal model (CSM) that is applicable in statistical frameworks by keeping the number of our vocoder parameters unchanged [26]. Experimental results from objective and subjective evaluations have shown that the proposed vocoder gives state-of-the-art vocoders performance in analysis-synthesis while outperforming the previous work of our continuous F0 based source-filter vocoder.

For the first time, we study in this paper the interaction between CSM and Bi-LSTM based RNN. We expect that the new model gives the performance of the high-quality WORLD vocoder based RNN.

The contribution of this paper is summarized as follows. 1) We propose a novel method to improve the accuracy of contF0 using time-warping approach, 2) We feed linguistic features to the Bi-LSTM based neural network to predict acoustic features, which are then passed to a CSM to generate the synthesized speech whilst at the same time being computationally efficient, 3) Through experiments, we provide detailed performance comparison based on objective evaluation and subjective listening test.

## III. PROPOSED METHODOLOGY

In this section, we describe our proposed model in three subsections. The first is dedicated to the time-warping approach that enhances the performance of the contF0, then the design of a continuous vocoder using sinusoidal synthesis algorithm will be introduced, and third the Bi-LSTM architecture and the training algorithm used in our experiment is described.

### A. Refining contF0 by Time-Warping approach

In speech signal, it is necessary that harmonic components are separated from each other with the purpose of being easily found and extracted. Once F0 rapidly changes, harmonic components are subject to overlapping each other and make it difficult to separate these components; or the close neighboring components make the separation through filtering very hard, especially with a low voice (such as male pitch) [33]. To overcome this problem, previous work in the literature has provided methods by introducing a time-warping based approach.

Abe et al. [34] incorporate time-warping into instantaneous frequency spectrogram frame by frame according to the change of the harmonic frequencies. In view of that, the observed F0 are seen to be constant within each analysis frame. More recently, time-warping pitch tracking algorithm have been also proposed by [35] which apparently had a significant positive impact on the voicing decision error and led to good results even in very noisy conditions. There has been another approach introduced by Stoter et al. [36] based on iteratively time-warping the speech signal and updating F0 estimate on time-warped speech, which has a nearly constant F0 over a segment of short duration that sometimes leads to inaccurate pitch estimates. To achieve a further reduction in the amount of contF0 trajectory deviation (deviate from their harmonic locations) and to avoid additional sideband components generation when a fast movement of higher frequencies occurs, adaptive time warping approach [37] combined with the instantaneous frequency can be used as a procedure to refine the contF0 algorithm.

We refer to the warping function as *p* which defines the relationship between two axes

$$\tau = p(t) \quad \text{and} \quad t = p^{-1}(\tau) \tag{1}$$

Where τ represents a time stretching factor. The first step is to stretch the time axis in order to make the observed contF0 value in the new temporal axis stay unchanged and preserves the harmonic structure intact [38] [39]. As the initial estimate of the contF0 is available, the second step of the refinement procedure is that the input waveform is filtered by bandpass filter bank $h(\tau)$ with different center frequencies $f_c$ multiplied by Nuttall window $w(\tau)$ [40] to separate only the fundamental component in the range near $f_c$

$$h(\tau) = w(\tau)\cos(2j\pi f_c \tau) \tag{2}$$

$$w(\tau) = 0.338946 + 0.481973 \cos\left(\frac{j\pi}{2} f_c \tau\right) \\ + 0.161054 \cos(j\pi f_c \tau) \\ + 0.018027 \cos\left(\frac{3j\pi}{2} f_c \tau\right) \tag{3}$$

Next, instantaneous frequencies $IF(\tau)$ of $h(\tau)$ have to be calculated. Flanagan's equation [41] is used to extract them from both the complex-valued signal and its derivative

$$IF_k(\tau) = \frac{a\frac{db}{d\tau} - b\frac{da}{d\tau}}{a^2 + b^2} \tag{4}$$

Where *a* and *b* are the real and imaginary parts of the spectrum of $h(\tau)$, respectively. *k* represents the harmonic number. As the $IF(\tau)$ indicates the value close to F0, the $\acute{contF0}$ is thus refined to a more accurate F0 by using a linear interpolation between $IF(\tau)$ values and contF0 coordinates. Then, using weighted average

$$\sum_{k=1}^{N} w_k \frac{\acute{ContF0}_k}{k} \tag{5}$$

Where $\sum_{k=1}^{N} w_k = 1$, provides a new $contF0_\tau$ estimate on the warped time axis. The last step is unwrapped in time to return the estimated value to the original time axis. Recursively applying these steps gives a final refined contF0 estimate.

With the purpose of assessing true performance of the refined contF0, a reference pitch contour (ground-truth) is required. Here, the ground truth is estimated from electro-glottal graph (EGG) as it is directly derived from glottal vibration and is largely unaffected by the nonharmonic components of speech [42].

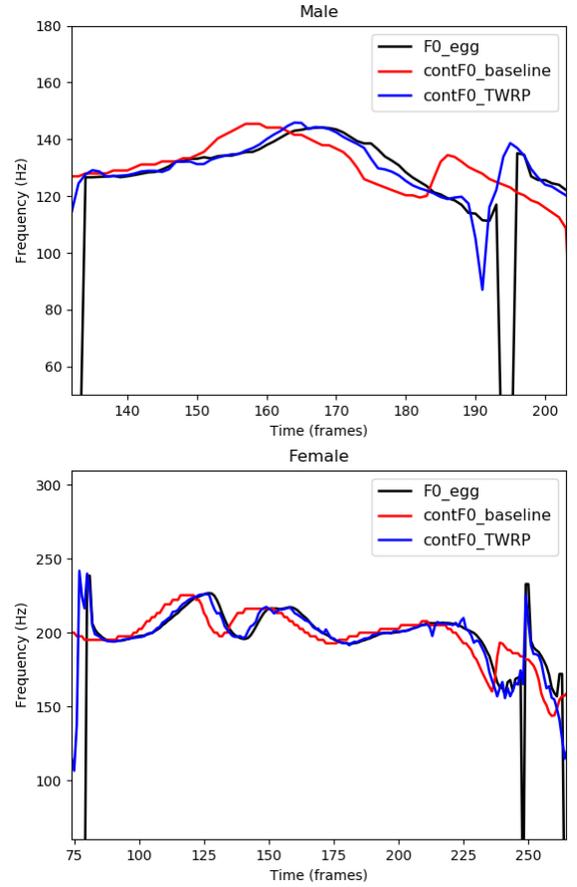

Fig. 1. Example of contF0 estimated by the time-warping (blue), baseline (red), plotted along with the ground-truth (black). Sentence: "Everything was working smoothly, better than I had expected.".

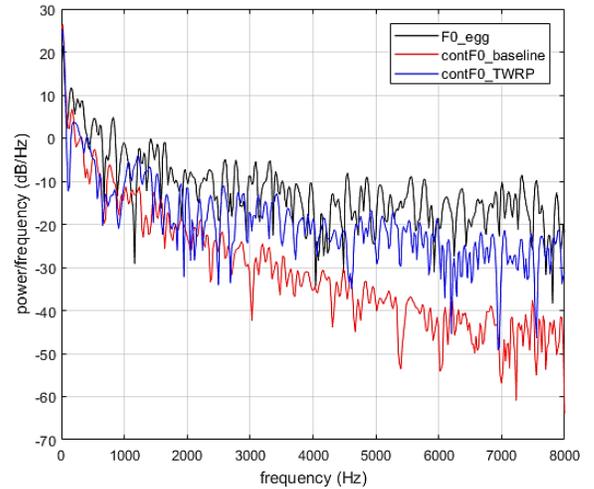

Fig. 2. The periodogram estimate of the PSD for the extracted contF0 trajectories.

An example of the proposed refinement on a male and female speech based on time-warping method is depicted in Fig. 1. Fig. 2 shows the power spectral density (PSD) calculated with the periodogram method for contF0 estimators.

Comparison to the ground-truth, it can be clearly seen that the contF0 trajectory given by time-warping method is a more accurate estimation than the baseline.

## B. Continuous Sinusoidal Model

Continuous vocoder based sinusoidal model (CSM) was designed to overcome shortcomings of discontinuity in the speech parameters and the computational complexity of modern vocoders. Moreover, the novelty behind this vocoder is to use harmonic features to facilitate and improve the synthesizing step before speech reconstruction.

By keeping the number of our previous source-filter vocoder parameters unchanged [26] and similarly to [29] [43], the synthesis algorithm implemented in this paper decomposes the speech frames into a lower-band voiced component $s_v(t)$ and an upper-band noise component $s_n(t)$ based on MVF values. We define these components here as

$$s(t) = s_v(t) + s_n(t) \quad (6)$$

In order to avoid discontinuities at the frames boundaries, Overlap-add (OLA) technique is used to reconstruct the speech signal from their corresponding parameters estimated from our analysis model in [26]. If the current frame is voiced, the harmonic part can be expressed as:

$$s_v^i(t) = \sum_{k=1}^{K^i} A_k^i(t) \cos\left(w_k^i t + \emptyset_k^i(t)\right) \quad (7)$$

$$w_k^i = 2\pi k (contF0)^i \quad (8)$$

Where $A_k(t)$ and $\emptyset_k(t)$ are the amplitude and phase at frame $i$ (both are obtained in a similar manner as described in [43]), $t = 0, 1, ..., N$ and $N$ is the frame length. $K$ is the time-varying frequency components or harmonics that depends on the contF0 and MVF as:

$$K^i = \begin{cases} round\left(\frac{MVF^i}{contF0^i}\right) - 1, & voiced\ frames \\ 0, & unvoiced\ frames \end{cases} \quad (9)$$

The synthetic noise signal $n(t)$ is filtered by a high-pass filter $f_h(t)$ with a cutoff frequency equal to the local MVF, and then modulated by its time-domain envelope $e(t)$ as we described it in our previous study [26]

$$s_n^i(t) = e^i(t) \left[f_h^i(t) * n^i(t)\right] \quad (10)$$

If the current frame is unvoiced, the harmonic part is zero and the synthetic frame is usually equal to the produced noise.

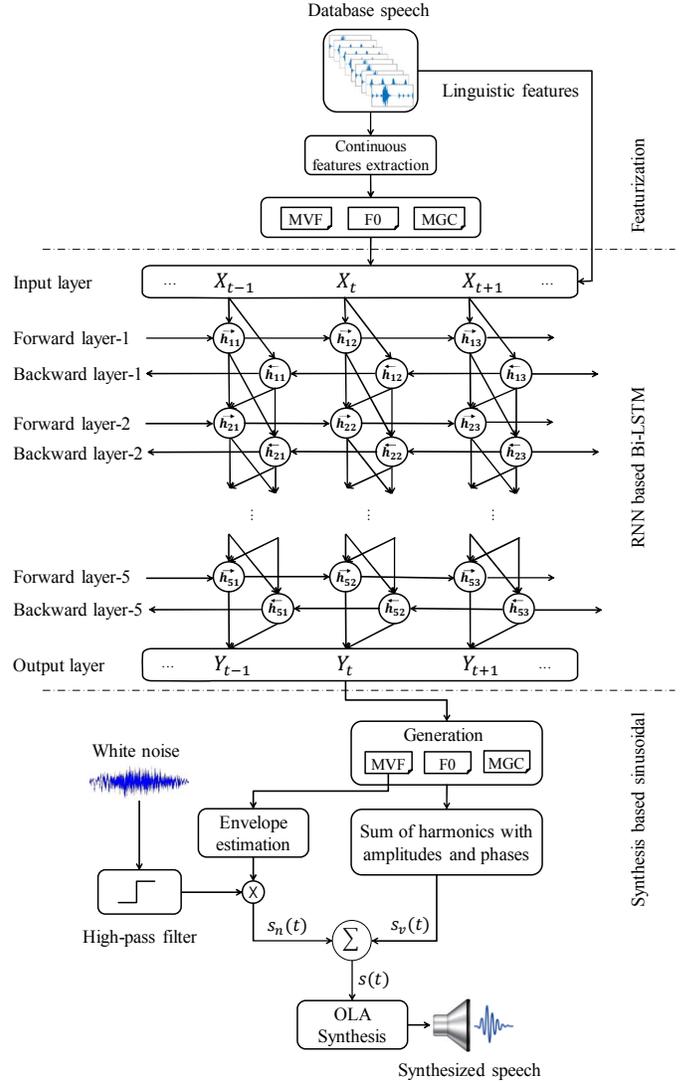

Fig. 3. Block diagram of the CSM based Bi-LSTM.

Hence, the synthesized speech signal is obtained by adding the harmonic and noise components. A block diagram of the proposed architecture is depicted in Fig. 3, where the bottom part is about CSM synthesis.

## C. Training model based on Bi-LSTM

The main concept of the Bi-LSTM was proposed in [17], and is a frequently used architecture for speech synthesis [18]. For a given input vector sequence $x = (x_1, ..., x_T)$, a regular RNN based Bi-LSTM calculates hidden state vector sequence $h = (h_1, ..., h_T)$ and outputs vector sequence $y = (y_1, ..., y_T)$. More specifically, Bi-LSTM separates the state neurons in a forward state sequence $\vec{h}$ (positive time direction), and backward state sequence $\overleftarrow{h}$ (negative time direction); which means that both forward and backward outputs are not connected. This can be observed in Fig. 3. The iterative process of the Bi-LSTM can be defined here as

$$\vec{h}_t = f(W_{x\vec{h}} x_t + W_{\vec{h}\vec{h}} \vec{h}_{t-1} + b_{\vec{h}}) \quad (11)$$

$$\overleftarrow{h}_t = f(W_{x\overleftarrow{h}} x_t + W_{\overleftarrow{h}\overleftarrow{h}} \overleftarrow{h}_{t-1} + b_{\overleftarrow{h}}) \quad (12)$$

$$y_t = W_{\vec{h}y} \vec{h}_t + W_{\overleftarrow{h}y} \overleftarrow{h}_t + b_y \quad (13)$$

Where $W$ is the connection weight matrix between two layers (e.g. $W_{xh}$ is the weight matrix between input and hidden vectors), $b$ is the bias vectors, and $f(\cdot)$ denotes an activation function which is defined as:

$$f(x) = \begin{cases} \dfrac{e^{2x} - 1}{e^{2x} + 1}, & \text{in the hidden layer} \\ \\ x, & \text{in the output layer} \end{cases} \quad (14)$$

Similarly to FF-DNN, RNN based Bi-LSTM aims to minimize the mean squared error function between the target output $y$ and the prediction output $\hat{y}$

$$E = \frac{1}{n} \sum_{i=1}^{n} (y_i - \hat{y}_i)^2 \quad (15)$$

By taking the advantages of FF-DNN and Bi-LSTM, a hybrid architecture can be used in this paper to firstly overcome the limitations discussed above and secondly to ensure an optimal training process can be used to enhance the performance of our vocoder. Consequently, 4 feed-forward hidden layers each consisting of 1024 units and performing a non-linear function of the previous layer's representation, followed by a single Bi-LSTM layer with 385 units, will be used in this work to train the CSM parameters.

## IV. EXPERIMENTAL SETUP

In order to evaluate the performance of the framework, a database containing a few hours of speech recorded carefully was required for giving analytical results. A reference system with a good quality performance is required to demonstrate the effectiveness and performance of the proposed methodology. Since the WORLD[1] vocoder [44] has a high-quality speech synthesis system for real-time applications and better than several high-quality vocoders [45], we use it as our state-of-the-art system based Bi-LSTM.

Datasets are described in more detail in the first part of this section, while training settings and error metrics are defined afterword.

### A. Dataset

The speech data used in this study consist of a database recorded for the purpose of developing text-to-speech (TTS) synthesis. Two English speakers were chosen from the CMU-ARCTIC[2] database [46], denoted AWB (Scottish English, male) and SLT (American English, female), which respectively consists of 1138 and 1132 sentences (roughly one hour for each speaker). The waveform sampling rate of the database is 16 kHz. In the experiments, 90% of these sentences were used for training and the rest were used for testing.

In the RNN-TTS experiments, 132 sentences from each speaker were analyzed and synthesized with the WORLD [44], baseline (that is our source-filter model [28]) and proposed vocoders. For WORLD and baseline vocoders, we used the same RNN architecture as for the proposed vocoder.

### B. Neural Network Setting

A hyperbolic tangent activation function was applied. The outputs lie in the range (-1 to 1) and this function can yield lower error rates and faster convergence than a logistic sigmoid function. For the first 15 epochs, a fixed learning rate of 0.002 was chosen with a momentum of 0.3. More specifically, after 10 epochs, the momentum was increased to 0.9 and then the learning rate was halved regularly. The Bi-LSTM used in this work was implemented in the open source Merlin toolkit for speech synthesis [47]. Besides, the training procedures were conducted on a high performance NVidia Titan X GPU. Weights and biases were prepared with small nonzero values, and optimized with stochastic gradient descent to minimize the mean squared error between its predictions and acoustic features of the training set.

### C. Error Metrics

A range of objective speech quality and intelligibility measures are considered to evaluate the quality of the proposed model. The results were averaged over the test utterances for each speaker. The following three evaluation metrics were used:

- **Log-Likelihood Ratio (LLR)** [48]**:** It is a distance measure that can be calculated from the linear prediction coefficients (LPC) vector of the natural and synthesized speech. The segmental LLR is

$$LLR = \frac{1}{N} \sum_{i=1}^{N} \log \left( \frac{a_{y,i}^T R_{x,i} a_{y,i}}{a_{x,i}^T R_{x,i} a_{x,i}} \right) \quad (16)$$

where $a_x$ and $a_y$ are the LPC all-pole gains of the natural and synthesized signal frames, respectively.

- **frequency-weighted segmental SNR (fwSNRseg)** [49]**:** $fwSNR_{seg}$ can be estimated by

---

[1] https://github.com/mmorise/World

[2] http://www.festvox.org/cmu_arctic/

$$fwSNR_{seg} = \frac{1}{N}\sum_{j=1}^{N}\left(\frac{\sum_{i=1}^{K} W_{i,j} \cdot \log \frac{X_{i,j}^2}{X_{i,j}^2 - Y_{i,j}^2}}{\sum_{i=1}^{K} W_{i,j}}\right) \quad (17)$$

where $X_{i,j}^2$, $Y_{i,j}^2$ are critical-band magnitude spectra in the $j^{th}$ frequency band of the natural and synthesized frame signals respectively, *K* is the number of bands, *W* is the weight vector defined in [50].

- **Log Spectral Distortion (LSD):** It can be defined as the square difference carried over the logarithm of the spectral envelopes of natural *X(f)* and synthesized *Y(f)* speech signals at *N* frequency points

$$LSD = \sqrt{\frac{1}{N}\sum_{i=1}^{N} mean\,(\log X(f_i) - \log Y(f_i))^2} \quad (18)$$

## V. EVALUATIONS AND RESULTS

After the development of the proposed methodology and experimental conditions, two main tests were conducted, aiming to launch consistent results.

### A. Objective Test

Here, we show the results for the error metrics presented in section IV. For all empirical measures, a calculation is done frame-by-frame, and a lower value indicates better performance except for the fwSNR$_{seg}$ (higher value is better). The results were averaged over 132 test sentences, and the best value in each column of Table 1 is bold faced.

It is good to note that the findings in Table I showed that the proposed vocoder based sinusoidal model succeeded in the Bi-LSTM training. Moreover, the CSM framework provides satisfactory results in terms of naturalness and intelligibility comparable to the high-quality WORLD and baseline that is the source-filter model. In particular, LLR between natural and synthesized speech frame is smaller than those using the baseline and WORLD methods. Focusing on the fwSNR$_{seg}$, it indicates that the WORLD model outperformed the proposed one only in the male speaker. While in terms of LSD, lowest correlation values were obtained with baseline method for all speakers. However, a slightly improvement was noted for the CSM over the WORLD model.

As a result, these experiments are showing that the proposed model with continuous sinusoidal model was beneficial in the statistical deep recurrent neural networks.

Table I. Average scores performance based on synthesized speech signal using proposed CSM for Male and Female speakers.

| Metrics | Model | AWB | SLT |
|---|---|---|---|
| LLR | Baseline | 1.4309 | 1.6966 |
| | Proposed | **1.4178** | **1.6791** |
| | WORLD | 1.5008 | 1.7516 |
| fwSNR$_{seg}$ | Baseline | 2.514 | 1.1882 |
| | Proposed | 2.4972 | **1.2278** |
| | WORLD | **2.5802** | 0.81389 |
| LSD | Baseline | **2.0739** | **2.2254** |
| | Proposed | 2.0995 | 2.2391 |
| | WORLD | 2.108 | 2.3373 |

### B. Subjective test

To demonstrate the efficiency of our proposed model, we performed a web-based MUSHRA (MUlti-Stimulus test with Hidden Reference and Anchor) listening test [51]. We compared natural sentences with the synthesized sentences from the baseline, proposed, WORLD, and an anchor system. The anchor type was the re-synthesis of the sentences with a standard pulse-noise excitation vocoder. In the test, the listeners had to rate the naturalness of each stimulus relative to the reference (which was the natural sentence), from 0 (highly unnatural) to 100 (highly natural). The utterances were presented in a randomized order (different for each participant). Altogether, 100 utterances were included in the test (2 speakers x 5 types x 10 sentences). 13 participants between the age of 24-38 (mean age: 31 years) were asked to conduct the online listening test. Five of them were males and eight were females. On average, each test was completed within 17 minutes. The listening test samples can be found online[3].

The results of the listening test are presented in Fig. 4 for the two speakers separately. For speaker AWB, it can be observed that the proposed framework significantly outperforms the baseline vocoder (Mann-Whitney-Wilcoxon ranksum test, with a 95% confidence level), while for speaker SLT, this difference is not significant. In both cases, the WORLD vocoder was rated slightly better than the continuous vocoder, but this difference is not significant. This means that CSM based RNN-TTS is close to the level of the state-of-the-art high quality vocoder than the baseline system.

---

[3] http://smartlab.tmit.bme.hu/ijcnn2019_vocoder

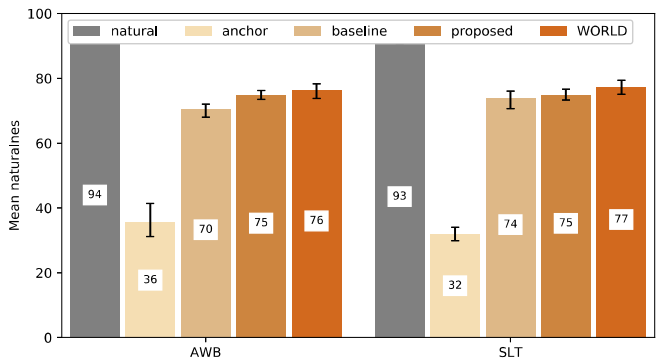

Fig. 4. MUSHRA scores for the naturalness question. Higher value means better naturalness. Errorbars show the bootstrapped 95% confidence intervals.

## VI. CONCLUSION

In this paper, we proposed a new approach to the statistical parametric speech synthesis using continuous sinusoidal model. The main idea is to integrate the CSM into Bi-LSTM deep recurrent neural network. The experiment was successful and we were able to add the continuous features (F0, Maximum Voiced Frequency, and Mel-Generalized Cepstrum) to the training framework based RNN.

Using a variety of measurements, the performance strengths and weaknesses of the proposed method for two different speakers were highlighted. From both objective and subjective evaluation metrics, the performance of the proposed system was superior to that of the baseline system. The proposed system was not found to be significantly different from the WORLD system, showing that it is close to the state-of-the-art.

For future work, the authors plan to apply the proposed sinusoidal model into voice conversion to further improve the perceptual quality of the converted speech.

ACKNOWLEDGMENT

The research was partly supported by the VUK (AAL-2014-1-183), by the DANSPLAT (EUREKA E!9944) projects and by the National Research, Development and Innovation Office of Hungary (FK 124584). The Titan X GPU used was donated by NVIDIA Corporation. We thank the subjects for participating in the listening test.

REFERENCES

[1] H. Zen, K. Tokuda, and A. Black., "Statistical parameteric speech synthesis," *Speech Communication,* vol. 51, no. 11, pp. 1039-1064, 2009.

[2] T. Yoshimura, K. Tokuda, T. Masuko, T. Kobayashi, and T. Kitamura, "Simultaneous modeling of spectrum, pitch, and duration in HMM based speech synthesis," *in Proceedings of Eurospeech,* pp. 2347-2350, 1999.

[3] K. Tokuda, T. Kobayashi, T. Masuko, T. Kobayashi, and T. Kitamura, "Speech parameter generation algorithms for HMM based speech synthesis," *in Proceedings of ICASSP,* pp. 1315-1318, 2000.

[4] Z. H. Ling et al., "Deep Learning for Acoustic Modeling in Parametric Speech Generation: A systematic review of existing techniques and future trends," *IEEE Signal Processing Magazine,* vol. 32, no. 3, pp. 35-52, 2015.

[5] H. Zen, K. Tokuda, and T. Kitamura, "Reformulating the HMM as a trajectory model by imposing explicit relationships between static and dynamic feature vector sequences," *Computer Speech Language,* vol. 21, no. 1, pp. 153-173, 2006.

[6] T. Toda, and K. Tokuda, "A speech parameter generation algorithm considering global variance for HMM-based speech synthesis," *IEICE Trans. Inf. Syst.,* Vols. E90-D, no. 5, pp. 816–824,, 2007.

[7] G. Hinton, and R. Salakhutdinov, "Reducing the dimensionality of data with neural networks," *Science,* vol. 313, no. 5786, pp. 504-507, 2006.

[8] O. Karaali, G. Corrigan, and I. Gerson, "Speech synthesis with neural networks," *in Proceedings of World Congress on Neural Networks,* pp. 45-50, 1996.

[9] Y. Xu, J. Du, L.R. Dai, and C.H. Lee, "An experimental study on speech enhancement based on deep neural networks," *IEEE Signal Processing Letter,* vol. 21, no. 1, pp. 65-68, 2014.

[10] S. Desai, E.V. Raghavendra, B. Yegnanarayana, A.W. Black, and K. Prahallad, "Voice conversion using artificial neural networks," *in Proceedings of ICASSP,* pp. 3893-3896, 2009.

[11] H. Zen, A. Senior, and M. Schuster, "Statistical parametric speech synthesis using deep neural networks," *in Proceedings of ICASSP,* p. 7962–7966, 2013.

[12] X. Yin, M. Lei, Y. Qian, F. K. Soong, L. He, Z.H. Ling, and L.R. Dai, "Modeling F0 trajectories in hierarchically structured deep neural networks," *Speech Communications,* vol. 76, pp. 82-92, 2016.

[13] Z. Wu, C. Valentini-Botinhao, O. Watts, and S. King, "Deep neural networks employing multi-task learning and stacked bottleneck features for speech synthesis," *in Proceedings of ICASSP,* pp. 4460 - 4464, 2015.

[14] S. Takaki, S. Kim, J. Yamagishi, and J.J. Kim, "Multiple feed-forward deep neural networks for statistical parametric speech synthesis," *in Proceedings of Interspeech,* pp. 2242-2246, 2015.

[15] Y. Bengio, P. Simard, and P. Frasconi, "Learning long-term dependencies with gradient descent is difficult," *in IEEE Transactions on Neural Networks,* vol. 5, no. 2, pp. 157-166, 1994.

[16] S. Hochreiter, and J. Schmidhuber, "Long short-term memory," *Neural computation,* vol. 9, no. 8, pp. 1735-1780, 1997.

[17] M. Schuster, and K. K. Paliwal, "Bidirectional recurrent neural networks," *IEEE Transactions on Signal Processing,* vol. 45, no. 11, pp. 2673-2681, 1997..

[18] Y. Fan, Y. Qian, F. Xie, and F. K. Soong, "TTS synthesis with bidirectional LSTM based recurrent neural networks," *in Proceedings of Interspeech,* pp. 1964-1968, 2014.

[19] A.V. Oord, S. Dieleman, H. Zen, K. Simonyan, O. Vinyals, A. Graves, N. Kalchbrenner, A.W. Senior, and K. Kavukcuoglu, "WaveNet: A Generative Model for Raw Audio," *in Proceedings of speech synthesis workshop (SSW),* 2016.

[20] S. Mehri, K. Kumar, I. Gulrajani, R. Kumar, S. Jain, J. Sotelo, A. Courville, and Y. Bengio, "SampleRNN: An unconditional end-to-end neural audio generation model," *International Conference on Learning Representations,* pp. 1-11, 2017.

[21] Y. Wang, et al, "Tacotron: towards end-to-end speech synthesis," *in Proceedings of Interspeech,* pp. 4006-4010, 2017.

[22] Q. Hu, K. Richmond, J. Yamagishi, and J. Latorre, "An experimental comparison of multiple vocoder types," in *in Proceedings of the ISCA SSW8,* pp.155–160, 2013.

[23] T.G. Csapó, G. Németh, and M. Cernak, "Residual-Based Excitation with Continuous F0 Modeling in HMM-Based Speech Synthesis," *3rd International Conference on Statistical Language and Speech Processing, SLSP 2015.,* vol. 9449, pp. 27-38, 2015.

[24] P. N. Garner, M. Cernak, and P. Motlicek, "A simple continuous pitch estimation algorithm," *IEEE Signal Processing Letters,* vol. 20, no. 1, pp. 102-105, 2013.

[25] T. Drugman, and Y. Stylianou, "Maximum Voiced Frequency Estimation: Exploiting Amplitude and Phase Spectra," *IEEE Signal Processing Letters,* vol. 21, no. 10, p. 1230–1234, 2014.

[26] M.S. Al-Radhi, T.G. Csapó, and G. Németh, "Time-domain envelope modulating the noise component of excitation in a continuous residual-based vocoder for statistical parametric speech synthesis," in *Proceedings of Interspeech,* Stockholm, pp. 434-438, 2017.

[27] M.S. Al-Radhi, T.G. Csapó, and G. Németh, "Continuous vocoder in feed-forward deep neural network based speech synthesis," in *Proceedings of digital speech and image processing*, Serbia, 2017.

[28] M.S. Al-Radhi, T.G. Csapó, and G. Németh, "Deep Recurrent Neural Networks in Speech Synthesis Using a Continuous Vocoder," *Speech and Computer (SPECOM), Lecture Notes in Computer Science,* pp. 282-291, 2017.

[29] Y. Stylianou, J. Laroche, and E. Moulines, "High-quality speech modification based on a harmonic + noise model," in *Proceedings of Eurospeech,* pp. 451-454, 1995.

[30] G. Degottex, and Y. Stylianou, "A Full-Band Adaptive Harmonic Representation of Speech," in *Proceedings of Interspeech,* pp. 382-385, 2012.

[31] Q. Hu, Y. Stylianou, R. Maia, K. Richmond, and J. Yamagishi, "Methods for applying dynamic sinusoidal models to statistical parametric speech synthesis," *IEEE ICASSP,* South Brisbane, QLD, pp. 4889-4893, 2015.

[32] M.S. Al-Radhi, T.G. Csapó, and G. Németh, "A Continuous Vocoder Using Sinusoidal Model for Statistical Parametric Speech Synthesis," *Speech and Computer (SPECOM), Lecture Notes in Computer Science,* vol. 1109, pp. 11-20, 2018.

[33] R. Kumaresan, and C.S. Ramalingam, "On separating voiced-speech into its components," in *Proceedings of the 27th Asilomar Conferance Signals, Systems, and Computers*, Pacific Grove, CA, pp. 1041-1046, 1993.

[34] T. Abe, T. Kobayashi, and S. Imai, "The IF spectrogram: a new spectral representation," in *Proceedings of the ASVA,* vol. 97, no. 1, pp. 423-430, 1997.

[35] S. Stone, P. Steiner, and P. Birkholz, "A time-warping pitch tracking algorithm considering fast f0 changes," in *Proceedings of the Interspeech,* Stockholm, Sweden, pp. 419-423, 2017.

[36] F.R. Stoter, N. Werner, S. Bayer, and B. Edler, "Refining Fundamental Frequency Estimates Using Time Warping," in *in Proceedings of the 23rd European Signal Processing Conference (EUSIPCO)*, Nice, France, 2015.

[37] H. Kawahara, Y. Agiomyrgiannakis, and H. Zen, "Using instantaneous frequency and aperiodicity detection to estimate f0 for high-quality speech synthesis," in *in Proceedings of the 9th ISCA Workshop on Speech Synthesis*, CA, USA, 2016.

[38] H. Kawahara, H. Katayose, A.D. Cheveigne, and R.D. Patterson, "Fixed point analysis of frequency to instantaneous frequency mapping for accurate estimation of f0 and periodicity," in *Proceedings of the EuroSpeech,* Budapest, pp. 2781-2784, 1999.

[39] N. Malyska, and T.F. Quatieri, "A time-warping framework for speech turbulence-noise component estimation during aperiodic phonation," in *Proceedings of the ICASSP,* Prague, Czech Republic, pp. 5404-5407, 2011.

[40] A. Nuttall, "Some windows with very good sidelobe behavior," *IEEE Trans. on acoust., speech, and signal processing,* vol. 29, no. 1, p. 84–91, 1981.

[41] J.L. Flanagan, and R.M. Golden, "Phase vocoder," *The Bell System Technical Journal,* vol. 45, no. 9, p. 1493–1509, 2009.

[42] J. Kominek, and A.W. Black, "CMU ARCTIC databases for speech synthesis," Carnegie Mellon University, 2003.

[43] D. Erro, I. Sainz, E. Navas, and I. Hernaez, "Harmonics Plus Noise Model Based Vocoder for Statistical Parametric Speech Synthesis," *IEEE Journal of Selected Topics in Signal Processing,* vol. 8, no. 2, pp. 184-194, 2014.

[44] M. Morise, F. Yokomori, and K. Ozawa, "WORLD: a vocoder-based high-quality speech synthesis system for real-time applications," *IEICE transactions on information and systems,* vol. 7, no. E99-D, pp. 1877-1884, 2016.

[45] M. Morise, and Y. Watanabe, "Sound quality comparison among high-quality vocoders by using re-synthesized speech," *Acoustical Science and Technology,* vol. 39, no. 3, pp. 263-265, 2018.

[46] J. Kominek, and A.W. Black, "CMU ARCTIC databases for speech synthesis," Carnegie Mellon University, 2003.

[47] W. Zhizheng, O. Watts, and S. King, "Merlin: An Open Source Neural Network Speech Synthesis System.," in *Proceedings of the 9th ISCA Speech Synthesis Workshop (SSW9),* Sunnyvale, CA, USA., 2016.

[48] S. Quackenbush, T. Barnwell, and M. Clements, Objective Measures of Speech Quality, Englewood Cliffs: NJ: Prentice-Hall, 1988.

[49] J. Ma, Y. Hu, and P. Loizou, "Objective measures for predicting speech intelligibility in noisy conditions based on new band-importance functions," *Acoustical Society of America,* vol. 125, no. 5, pp. 3387-3405, 2009.

[50] ANSI S3.5-1997, "Methods for the calculation of the speech intelligibility index," 1997.

[51] ITU-R Recommendation BS.1534, "Method for the subjective assessment of intermediate audio quality," 2001.